\documentclass[12pt]{iopart}

\usepackage{graphicx}
\usepackage{color}

\begin{document}

\title[Inner products of resonances]{Inner products of resonance solutions in 1-D quantum barriers}

\author{J. Julve}

\address{IFF, Consejo Superior de Investigaciones
Cient\'\i ficas, Serrano 113 bis, Madrid 28006, Spain}

\ead{julve@imaff.cfmac.csic.es}

\author{F. J. de Urr\'{\i}es}

\address{Departamento de F\'\i sica,
Universidad de Alcal\'a de Henares, Alcal\'a de Henares (Madrid),
Spain}

\ead{fernando.urries@uah.es}

\begin{abstract}
The properties of a prescription for the inner products of the
resonance (Gamow states), scattering (Dirac kets), and bound states
for 1-dimensional quantum barriers are worked out. The divergent
asypmtotic behaviour of the Gamow states is regularized using a
Gaussian convergence factor first introduced by Zel'dovich. With
this prescription, most of these states (with discrete complex
energies) are found to be orthogonal to each other and to the Dirac
kets, except when they are neighbors, in which case the inner
product is divergent. Therefore, as it happens for the continuum
scattering states, the norm of the resonant ones remains
non-calculable. Thus, they exhibit properties half way between the
(continuum real) Dirac-$\delta$ orthogonality and the (discrete
real) Kronecker-$\delta$ orthogonality of the bound states.

\pacs{03.65.Nk}

\submitto{Journal of Physics A: Mathematical and Theoretical}

\end{abstract}

\maketitle

\section{Introduction}

Resonances in Quantum Mechanics describe states evolving
non-unitarily \cite{Nussenzveig}, both decaying and building ones,
and have found applications in the study of nuclear reactions. As
such, they have been more often studied in the case of spherical
nuclear potentials, mainly of the "shell-model" type and restricting
the analysis to the radial s-wave equation. Albeit for obvious
unessential differences in the Boundary Conditions, the physical
picture is similar to the case of the 1-dimensional barriers we are
considering, more akin to condensed matter systems or ion-trapping
devices. The interest is alive and work is still in progress on the
physical interpretation of resonances \cite{Hatano}.

Resonant solutions to the $\rm Schr\ddot{o}dinger$ equation occur in
any simple potential barrier with a compact support, the plain
square barrier being the most simple, tractable and fully
representative example, and they are more often found in alternative
specialized expansions of the Green Function and other physically
relevant objects. They correspond to complex energies and momenta,
which place them out of the familiar realm of Hermitian operators in
Hilbert spaces, where a wealth of well known mathematical properties
is available. Though having been first studied in early times
\cite{Gamow}\cite{Siegert}, the non-trivial mathematical properties
of resonances have spurred a long lasting investigation effort. This
paper focuses on some still debatable issues regarding the norm,
inner products and completeness properties of resonant states.

The issue of the completeness arises in its simplest form as soon as
one attempts, for instance, to expand the identity in terms of
projectors on bound (if any) $|\phi_i\rangle$ , resonance
$|z_n\rangle$ and background ("scattering" complex energy)
$|z\rangle$ states, namely
\begin{equation}
I=\sum_i|\phi_i\rangle\langle\phi_i|+\sum_n|z_n\rangle\langle
z_n|+\oint\rmd z\;|z\rangle\langle z|\;\;.
\end{equation}
where the second sum may involve a variable number and type of
resonances. Such an expansion is, in principle, attainable by
deforming, in the complex (two-sheet) plane \cite{Berggren1}, the
continuum real energy integration occurring in the traditional
expansion in terms of bound and scattering states
\begin{equation}
I=\sum_i|\phi_i\rangle\langle\phi_i|+\int\rmd E\;|E\rangle\langle E|
\end{equation}
following the lines of the proof of (2), as given for instance by
\cite{Newton}. In particular cases a complete expansion in terms of
resonances (plus bound states) can be found, as for example for
continuum wave functions (or the scattering solutions) {\it within}
a finite region including the support of the barrier
\cite{Calderon3} \cite{Calderon4} \cite{Julve}. This has direct
application for instance in time-dependent problems, as in the time
evolution of quantum decay \cite{Calderon5}. In the general case,
covering the whole space, directly testing any resonance expansion
like (1) against the idempotence requirement $I^2=I$, needs the
computation of the inner products of all these families of states
with each other. We aim to perform some steps in this direction.

The difficulty stems from the divergent asymptotic behaviour of the
resonance solutions, which leads to infinite norms and seemingly
divergent and hard to calculate inner products, and has given rise
to a variety of proposals to circumvent it. These proposals adopt
different prescriptions to render finite the space integrals
involved in these products and in general matrix elements. We may
quote analytical continuations of the resonant solutions in the
complex momentum plane \cite{Romo1}, the "External Complex Scaling"
\cite{Zavin} of the space-coordinate integration variable, and the
introduction of convergence factors in the integrals.

In this paper we shall adopt a Gaussian convergence factor, first
introduced by Zel'dovich \cite{Zel'dovich} and used by others
\cite{Berggren1}, with which a limit can be worked out yielding
novel non-trivial results: in fact we recover the result of the
integrals that were already well defined, and extend the finite
result to a newer region of the complex momentum plane.  The
procedure, which avoids relying on analytical continuation
arguments, leads to a prescription for the definition of the inner
products, yielding a specific set of orthogonality relations. Our
results are derived for a general potential with finite support of
which the bound and resonant poles of the S-matrix are known.

The usual inner product in the Hilbert space, involving the complex
conjugate of one of the wave functions, is defined so as to have
real probability densities and norms for the general wave functions.
However when dealing with resonant solutions, a symmetrical inner
product is involved in the convolution of the Green Function with
initial states (superpositions of scattering states), and is
associated to a complex "norm" \cite{Calderon2}, so we explore the
properties of both alternative definitions.

The plan of the paper is the following: In Section 2 we briefly
review the relevant features of the resonances while fixing some
notation and stressing the relationships between proper, anti,
outgoing and incoming resonant solutions. In Section 3 we calculate
the standard inner product of resonant states between themselves and
with the bound and the scattering states. In Section 4 we introduce
the symmetrical inner product, and give the result for the
resonant-resonant and resonant-scattering product. Then in Section 5
we outline in some detail the regularization prescription adopted,
based on a Gaussian factor, and show how the (either finite or
divergent) results are recovered in the non-regularized limit. For
comparison, we comment here on some other prescriptions found in the
literature. Finally the Conclusions are drawn in Section 6. The
crucial integral formulas are derived in an Appendix.

\section{The resonant solutions}

We consider the 1-D time-independent $\rm Schr\ddot{o}dinger$
equation
\begin{equation}
[\frac{\partial^2}{\partial x^2}+p^2-2mV(x)]\;\psi(x,p) =0
\end{equation}
where we use units such that $\hbar=1$.

For a cut-off potential $V(x)$ describing a general barrier with
support in the compact interval [0,\emph{L}], besides the usual
scattering \emph{in} and \emph{out} solutions for continuous real
energy $E=p^2/2m\,>0$ and possible bound states (discrete real
$E_i<0)$, one has resonant solutions (Gamow states) satisfying the
homogeneous \emph{Outgoing} Resonant Boundary Conditions (ORBCs),
\begin{equation}
\partial_x \psi\!\!\mid_{x=0}\,=-\rmi p\,\psi(0) \;\;,\;\;
\partial_x \psi\!\!\mid_{x=L}\,=\rmi p\,\psi(L)\;.
\end{equation}
Solutions $u_n(x)$, called proper (\emph{Outgoing}) resonances,
exist for a denumerable set of isolated values $p_n$ of $p\;$ (with
corresponding energies $z_n=p^2_n/2m$) lying inside the octants
close to the real axis (i.e. $|{\rm Re}\,p_n|>|{\rm Im}\,p_n|$) in
the lower half complex plane (i.e. ${\rm Im}\,p_n<0$), and occupy
symmetrical positions with respect to the imaginary axis (see Figure
1). It is customary to label them as $p_n$ ($n=1,2,...$) when ${\rm
Re}\,p_n
>0$ and as $p_{-n}\equiv-p_n^*$ their symmetric ones, some times
called anti-resonances. In the case of a simple square barrier, the
real parts ${\rm Re}\,p_n$ tend to be spaced regularly for
increasing $|n|$, while $|\,{\rm Im}\,p_n|$ grows slowly
\cite{Nussenzveig}.

RBCs with reversed sign of $p$ in (4) correspond to \emph{Incoming}
solutions $\tilde{u}_ n(x)$, and the momenta
$\tilde{p}_n\equiv-{p}_n=p^*_{-n}$ lie in the upper half complex
plane. We denote with $\tilde{u}_ n(x)=u_n(\tilde{p}_n;x)$ these
solutions and with $|\tilde{z}_n\rangle$ the corresponding states.
Notice that the corresponding energies $\tilde{z}_n$ lie in the
first Riemann sheet. For real potentials $V(x)$, the complex
conjugates $u^*_n(x)$ of the outgoing resonances correspond instead
to yet outgoing solutions with mirror momenta $-p_n^*$ , and
energies $z^*_n$ in the second Riemann sheet, so that $u^*_n(x)=
u_{-n}(x)$ and $z^*_n=z_{-n}$. For the same reason, the complex
conjugates of the incoming resonances are again incoming solutions.
In the literature it has been easily mistaken $u^*_n(x)$ for an
incoming solution.

The ORBCs are equivalent to imposing the asymptotic form
\begin{equation}u_n(x)=\left\{
\begin{array}{ll}
R_n\,e^{-{\rm
i}p_nx}&\quad,\quad x\leq 0\\
T_n\,e^{{\rm i}p_nx}&\quad,\quad x\geq L
\end{array}\right.
\end{equation}
where the amplitudes $R_n$ and $T_n$ differ by a phase and are
defined up to a global arbitrary factor. An immediate consequence is
that the norm
\begin{equation}
\|u_n\|\equiv\langle z_n|z_n\rangle=\int^{+\infty}_{-\infty}\rmd
x\;u^*_n(x)u_n(x)\approx\int^{+\infty}_{-\infty}\rmd x \exp{2|x||{\rm
Im}\, p_n| } =\infty
\end{equation}
is even more divergent than for the Dirac states
$\|\psi\|\equiv\langle
E|E\rangle=\delta(0)\approx\int^{+\infty}_{-\infty}\rmd x=\infty$,
causing both kinds of states not to belong to ${\cal L}^2$.

This fact, as opposite to the finite norm of the bound states, has
been addressed  in the framework of Rigged Hilbert Spaces (RHS)
\cite{Bohm}\cite{Madrid1} along the same lines adopted for the Dirac
kets. In RHS, these states are interpreted just as linear
functionals on more restricted spaces (Hilbert or Schwartz) in which
an inner product is properly defined, but nevertheless it is
customary to consider, in a somewhat relaxed sense, inner products
of the Dirac kets which have a meaning only as distributions, namely
$\langle E|E'\rangle=\delta(E-E')$. However, when one tries to give
an answer of this type for the resonance states, the inner products
\begin{equation}
\langle z_n|z_m\rangle=\int^{+\infty}_{-\infty}\rmd
x\;u^*_n(x)u_m(x)
\end{equation}
may be expected to be generally divergent as well, because of the
exponential growth for large $x$ . For this reason their actual
calculation has spurred the adoption of a number of elaborated
strategies.

The same difficulties affect the inner products $\langle
z_n|E\rangle$ between resonant and scattering states and, more
generally, between resonant states and wave packets. This is an
important issue as long as the analytical calculation of, for
instance, the time evolution of an initial state impinging on a
barrier \cite{Julve}, or of the shutter problem \cite{Calderon1},
contains sums of terms with poles in the resonant momenta. For the
(either Laplace or Fourier) transform of the time dependent Green
Function one has \cite{Calderon2}
\begin{equation}
G(x,x';p)=\sum_n\frac{1}{N_n}\frac{1}{2p_n}\frac{u_n(x)u_n(x')}{p-p_n}+
\rm{regular\,terms}
\end{equation}
The interpretation of the pole terms as (transient) excitations of
the resonant eigenmodes of the system requires the knowledge of the
projection of the initial state on the resonant ones. The
dimensionless normalization factor in (8) is
\cite{Calderon2}\cite{Calderon1}
\begin{equation}
N_n=\rmi\frac{u^2_n(a)+u^2_n(b)}{2p_n}+\int^b_a\rmd x\;u^2_n(x)\;,
\end{equation}
where $a=0$ and $b=L$, which is a sort of (complex) "norm". Notice
that, when applied to wave functions $\psi(x)\in {\cal L}^2$, it
reminds the usual (real) norm in Hilbert space albeit for replacing
the squares by the square modulus and letting $a$ and $b$ recede
respectively to $-\infty$ and to $+\infty$. For the resonant
solutions however, $N_n$ diverges in this limit.

\section{Inner products}

In this section we shall calculate the usual inner products $\langle
z_n|z_m\rangle$ of resonant states with themselves (7) and their
mixed products $\langle z_n|E\rangle$ and $\langle\phi_i
|z_n\rangle$ with the scattering and bound states respectively. The
method leads also to the usual inner products $\langle\phi_i
|\phi_j\rangle=\delta_{ij}$ , $\langle\phi_i |E\rangle=0$ , and
$\langle E|E'\rangle=\delta(E-E')$.

\subsection{Resonances}

The space integral giving the inner product of two resonant
solutions can be split in three sectors
\begin{equation}
\fl\begin{array} {ll}\langle z_n|z_m\rangle
&\equiv\int^{+\infty}_{-\infty}\rmd x\,u_n^*(x)u_m(x)\\
&=R^*_nR_m\int^{0}_{-\infty}\rmd
x\,e^{\rmi(p^*_n-p_m)x}+\int^L_0\rmd
x\,u^*_nu_m+T^*_nT_m\int^{\infty}_L\rmd
x\,e^{-\rmi(p^*_n-p_m)x}\\
&=(R^*_nR_m+T^*_nT_m)\int_0^{\infty}\rmd
x\,e^{-\rmi(p^*_n-p_m)x}+\int^L_0\rmd
x\,u^*_nu_m-T^*_nT_m\int^L_0\rmd x\,e^{-\rmi(p^*_n-p_m)x}
\end{array}
\end{equation}

The ORBCs in $0$ and $L$, together with their complex conjugates,
allow to express the finite integral $\int^L_0\rmd x\,u^*_nu_m$ in
terms of the amplitudes $R$ and $T$ outside the barrier, regardless
of the explicit form of $u_n(x)$ inside the barrier and hence of the
particular form of the potential. The procedure uses the $\rm
Schr\ddot{o}dinger$ operator $O\equiv-2mH=\partial^2_x-2mV$ and
integration by parts to obtain
\begin{eqnarray}
\fl(p^{*2}_n-p^2_m)\int^L_0\rmd
x\,u^*_nu_m&=&2m(z^*_n-z_m)\int^L_0\rmd x\,u^*_nu_m \\\nonumber
&=&\int^L_0\rmd
x\,u_n^*(x)(\overrightarrow{O}-\overleftarrow{O})u_m(x)=W[ u^*_n,
u_m]^L_0\\\nonumber
\end{eqnarray}
where $W[\phi\,,\psi]\equiv
\phi\,\partial_x\psi-\psi\partial_x\,\phi$ is the Wronskian of the
functions $\phi$ and $\psi$ , so that
\begin{equation}\int^L_0\rmd
x\,u^*_nu_m=\frac{\rmi}{p^*_n-p_m}(T^*_nT_me^{-\rmi(p^*_n-p_m)L}+R^*_nR_m)
\end{equation}
Then cancelations occur such that
\begin{equation}\langle
z_n|z_m\rangle=(R^*_nR_m+T^*_nT_m)(\int_0^{\infty}\rmd
x\,e^{-\rmi(p^*_n-p_m)x}+\frac{\rmi}{p^*_n-p_m})
\end{equation}

Now the point is the calculation of the integral in (13). We adopt
the limit $\lambda\rightarrow 0$ in (29) as a prescription for the
result of (13), and defer the discussion of our scheme to Section 5
and to the Appendix.

With this prescription we finally obtain
\begin{equation}
\langle z_n|z_m\rangle= \left\{
\begin{array}{ll}
 0&\quad,\quad-\frac{\pi}{4}< {\rm arg}(p_m-p^*_n)<5\frac{\pi}{4}\\
\infty&\quad,\quad {\rm otherwise}
\end{array}\right.
\end{equation}
In particular $\langle z_n|z_n\rangle= \infty$, as expected.

For each resonant state $|z_n\rangle$, the result (14) defines a
"neighborhood of divergence" so that $|z_n\rangle$ is orthogonal to
any other $|z_m\rangle$ the momentum $p_m$ of which lies outside a
"cone of divergence" with apex in $p^*_n$ , and gives a divergent
inner product if $p_m$ lies inside this cone (Figure 1)

\begin{figure}[h]
\begin{center}
\includegraphics[width=0.8\textwidth]{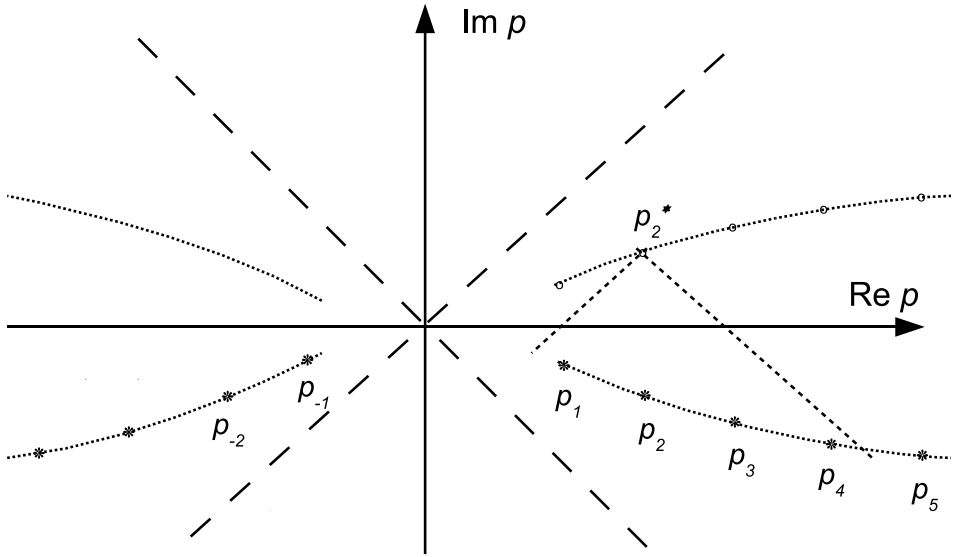}
\caption{An example of orthogonality between resonant
states:\hfill\break $\langle z_2|z_n\rangle=0$ for $n<0$ and $n>4$ ,
while $\langle z_2|z_m\rangle=\infty$ for $0<m\leq4$ .}
\label{Fig.1}
\end{center}
\end{figure}

\subsection{Resonant and Scattering states}

The \emph{in} and \emph{out} scattering solutions obey a single
differential BC at one of the points $x=0$ or $x=L$ . For instance,
a right-moving \emph{in} state obeys only the second BC in (4) with
$p>0$. This is equivalent to imposing the asymptotic form
\begin{equation}
\psi^+_r(x)= \left\{
\begin{array}{ll}
\rme^{{\rm i}px}+R(p)\,\rme^{-{\rmi}px}&\quad,\quad x\leq 0\\
T(p)\,\rme^{{\rm i}px}&\quad,\quad x\geq L
\end{array}\right.
\end{equation}
For instance, the combined BCs of the resonant and of the ($p>0$ ,
\emph{in}) scattering solutions similarly lead to
\begin{eqnarray}\langle
z_n|E\rangle=&\int_0^{\infty}\rmd
x\,(R^*_n\,e^{-\rmi(p^*_n+p)x}+(R^*_nR+T^*_nT)\,e^{-\rmi(p^*_n-p)x})\\\nonumber
&+\frac{\rmi}{p^*_n-p}(R^*_nR+T^*_nT)+\frac{\rmi}{p^*_n+p}R^*_n\\\nonumber
\end{eqnarray}
With $p>0$ , for $n>0$ we always have $-\frac{\pi}{4}< {\rm
arg}(p^*_n+p)<5\frac{\pi}{4}$ , so that
\begin{equation}\langle
z_n|E\rangle=(R^*_nR+T^*_nT)(\int_0^{\infty}\rmd
x\,e^{-\rmi(p^*_n-p)x} +\frac{\rmi}{p^*_n-p})
\end{equation}
and therefore, with the prescription adopted,
\begin{equation}
\langle z_n|E\rangle= \left\{
\begin{array}{ll}
 0&\quad,\quad-\frac{\pi}{4}< {\rm arg}(p-p^*_n)<5\frac{\pi}{4}\\
\infty & \quad,\quad{\rm otherwise}
\end{array}\right.
\end{equation}
This means that a given scattering \emph{in} state $|E\rangle$ (with
momentum $p>0$ on the real axis) is orthogonal to any $|z_n\rangle$
(with $n>0$) if the momentum $p_n$ lies outside the cone with apex
in $p$ , the inner product being divergent otherwise. Viceversa,
given $p_n$ , the momenta $p\,$ of the orthogonal scattering states
lie outside the cone with apex in $p_n$ (Figure 2).

\begin{figure}[h]
\begin{center}
\includegraphics[width=0.8\textwidth]{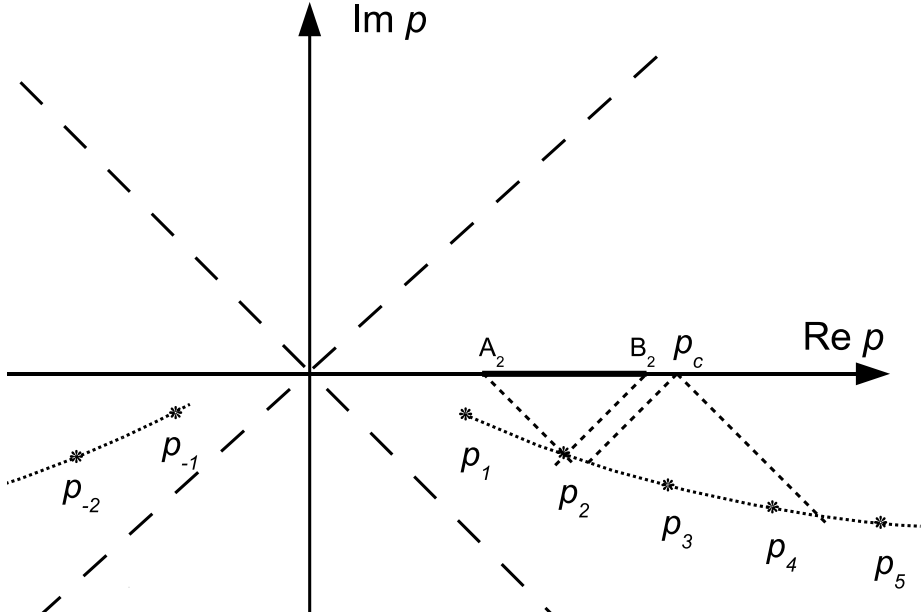}
\caption{Two examples of orthogonality between resonant and
scattering states: $\langle z_2|E\rangle=0\;(\infty)$ when
$p\equiv\sqrt{2mE}\subset \textsf{R}_+$ lies out of (inside) the
interval $[A_2,B_2]$ , \break and $\langle
E_c|z_n\rangle=0\;(\infty)$ when $n\neq3,4$ ($n=3,4$) respectively.}
\label{Fig.2}
\end{center}
\end{figure}

The scattering states always have a reflected wave with momentum of
opposite sign to the incident one, so the situation is trickier for
$n<0$ . Given $p_{-|n|}$ , the cone to be considered is again the
one with apex in the mirror momentum $p_{\,|n|}$ .

\subsection{Resonant and bound states}

Let us suppose that, besides the barrier, there is some potential
well within the region $[0,L]$ sustaining bound states
$|\phi_i\rangle$ with purely imaginary momenta $p_i=\rmi
q_i\;(q_i>0)$. The exponential decrease of the amplitude outside the
well, corresponding to resonant-like BCs
\begin{equation}
\partial_x \phi\!\!\mid_{x=0}\,=q\,\phi(0) \;\;,\;\;
\partial_x \phi\!\!\mid_{x=L}\,=-q\,\phi(L)\;\;\;\;\;\;\;\;\;\;(q>0)\;,
\end{equation}
which imply the more popular (and weaker) $\;\phi_i(\pm\infty)=0$,
may be translated into the assumption of an asymptotic form similar
to (5), namely
\begin{equation}\phi_i(x)=\left\{
\begin{array}{ll}
R_i\,e^{q_ix}&\quad,\quad x\leq 0\\
T_i\,e^{-q_ix}&\quad,\quad x\geq L
\end{array}\right.
\end{equation}
which is manifestly square integrable.

Following the same steps leading to (13) we now have
\begin{equation}\langle
\phi_i|z_n\rangle=(R^*_iR_n+T^*_iT_n)(\int_0^{\infty}\rmd
x\,e^{\rmi(\rmi q_i+p_n)x}-\frac{\rmi}{\rmi q_i+p_n})
\end{equation}
The integral is convergent, yielding $\,\rmi (\rmi q_i+p_n)^{-1}$
and hence $\langle \phi_i|z_n\rangle=0$ , provided that
$q_i>|\,{\rm\,Im}\,p_n |$ . Thus we have a situation similar to that
occurring between the resonances, namely that the states are
orthogonal if the bound state momentum, lying in the positive
imaginary axis, and the resonant one, lie outside the respective
divergence cones, the inner product being infinite otherwise. It
might happen that, for general analyticity reasons for any general
potential $V(x)$ with barriers and wells within $[0,L]$, the cones
of the allowed bound state momenta do never include the resonant
momenta, but this point would require separate investigation.

\section{Symmetrical inner products}

Together with the standard definition (7), which yields (finite or
infinite) real norms, we shall consider also the alternative
symmetrical definition
\begin{equation}
\{\Phi|\Psi\}\equiv\int^{+\infty}_{-\infty}\rmd
x\;\phi(x)\psi(x)\;\;\;,
\end{equation}
which arises in the convolution of the kernel (8) with the initial
state, and yields complex "norms".

The calculation follows the same lines above relying on the Boundary
Conditions:
\begin{equation}\{z_n|z_m\}=(R_nR_m+T_nT_m)(\int_0^{\infty}\rmd
x\,e^{\rmi(p_n+p_m)x}-\frac{\rmi}{p_n+p_m})
\end{equation}
The result is
\begin{equation}
\{z_n|z_m\}=\left\{
\begin{array}{ll}
0&\quad ,\quad-\frac{\pi}{4}< {\rm arg}(p_m+p_n)<5\frac{\pi}{4}\\
\infty &\quad ,\quad{\rm otherwise}
\end{array}\right.
\end{equation}
For a given $|z_n\rangle$, the apex of the divergence cone is at
$-p_n$ so that, in particular, $\{z_n|z_m\}= 0$ if both $m,n>0$,
also if both $m,n<0$, and, more noteworthy,  $\{z_n|z_n\}= 0$ for
any $n$ . Only some of the products are divergent when $n$ and $m$
have opposite sign.

For the crossed products of $|z_n\rangle$ ($n>0$) with the
scattering $p>0\;$ \emph{in} state one has
\begin{eqnarray}\{
z_n|E\}=&\int_0^{\infty}\rmd
x\,(R_n\,e^{\rmi(p_n-p)x}+(R_nR+T_nT)\,e^{\rmi(p_n+p)x})\\\nonumber
&-\frac{\rmi}{p_n+p}(R_nR+T_nT)-\frac{\rmi}{p_n-p}R_n\\\nonumber
\end{eqnarray}
One always has $-\frac{\pi}{4}< {\rm arg}(p_n+p)<5\frac{\pi}{4}$ ,
so that
\begin{equation}
\{z_n|E\}=\left\{
\begin{array}{ll}
0& \quad ,\quad -\frac{\pi}{4}< {\rm arg}(p_n-p)<5\frac{\pi}{4}\\
\infty & \quad ,\quad {\rm otherwise}
\end{array}\right.
\end{equation}
Therefore the divergence cone has the apex in $p_n$ ,
as in the case of the standard inner product $\langle z_n|E\rangle$.
For $n<0$ , only $p+p_{-|n|}$ may lie in the sector
$\,5\frac{\pi}{4}< {\rm arg}(p_n+p)<7\frac{\pi}{4}\;$ leading to a
divergent product, so that, for a given $p_{-|n|}$ the $p\,$-states
are orthogonal if $p$ lies inside the cone with apex in the mirror
momentum $p_{\,|n|}$, reproducing the same situation encountered for
the standard product.

Because of the relationship ${u}_{-n}(x)=u^*_n(x)$ between the
outgoing solutions, we have $\langle z_n|z_m\rangle=\{z_{-n}|z_m\}$
and $\{z_n|z_m\}=\langle z_{-n}|z_m\rangle$. Then the result
$\{z_n|z_n\}= 0$ is less surprising if rewrite it as $\langle
z_{-n}|z_n\rangle=0$ , just a particular case of the conventional
orthogonality between states with momenta in opposite quadrants of
the lower half complex plane.

\section{Regularization of the divergent products}

Several proposals have been worked out in the literature to deal
with the divergent inner products. We quote here the one by W. Romo
\cite{Romo1} of analytically continuing the momentum dependence of
the solutions $u(x)$ inside the products from the upper-half complex
plane, where the (outgoing) function would be square-integrable, to
the lower half-plane where the resonant momenta lie. This is
equivalent to \emph{prescribe} the finite value $\rmi k^{-1}$ for
the result of the integral $\int^{\infty}_0\rmd x \,e^{\rmi kx}$
also in the whole lower half-plane $k$, where it would actually be
divergent (see the discussion in Appendix). This approach, combined
with advantages of working in momentum representation, has been
followed in later works \cite{Mondragon}.

Other proposals rely on the introduction of convergence factors in
the integral which are able to cope with the asymptotic exponential
growth of the resonant solutions. The simple factor $e^{-\lambda
x},\,(\lambda\; {\rm real}>0)$ works only if $\lambda$ is greater
than the absolute values of the imaginary parts of given $p_n$ and
$p_m$ (which grow with increasing $n\,,m$), so that $\lambda$ cannot
be brought to zero without rendering divergent the integral. However
it lets performing an analytic continuation in $\lambda$ to its
"forbidden" values which may be adopted as a prescription.

Following Zel'dovich \cite{Zel'dovich}, we adopt the more powerful
Gaussian factor $e^{-\lambda x^2}$, which is able to overcome the
growth of all the resonant solutions even for $\lambda\rightarrow
0_+$ , and is computationally tractable
\cite{Zel'dovich}\cite{Berggren1}:
\begin{equation}
\langle\!\langle
z_n|z_m\rangle\!\rangle_{\lambda}\equiv\int^{+\infty}_{-\infty}\rmd
x\;e^{-\lambda x^2}u^*_n(x)u_m(x)\;\;\;\;(\lambda\,\,{\rm real}
>0)\end{equation}
The functions $e^{-\frac{1}{2}\lambda x^2}u_n(x)$ are not
eigenfunctions of the Hamiltonian so the exact calculation of (27)
requires the knowledge of the explicit form of the solutions inside
the barrier.

Equation (10) now reads
\begin{eqnarray}
\langle\!\langle
z_n|z_m\rangle\!\rangle_{\lambda}&=&(R^*_nR_m+T^*_nT_m)\int_0^{\infty}\rmd
x\,e^{-\lambda x^2}e^{-\rmi(p^*_n-p_m)x}\\\nonumber &&+\int^L_0\rmd
x\,e^{-\lambda x^2}u^*_nu_m-T^*_nT_m\int^L_0\rmd x\,e^{-\lambda
x^2}e^{-\rmi(p^*_n-p_m)x}\\\nonumber
\end{eqnarray}
According to (A.1), this regularized expression is finite for
$\lambda > 0$ :
\begin{eqnarray}
\langle\!\langle
z_n|z_m\rangle\!\rangle_{\lambda}&=&(R^*_nR_m+T^*_nT_m)\frac{-\rmi}{p^*_n-p_m}\sqrt{\pi}\,z\,e^{z^2}\,{\rm
erfc}(z)\\\nonumber &&+\int^L_0\rmd x\,u^*_nu_m-T^*_nT_m\int^L_0\rmd
x\,e^{-\rmi(p^*_n-p_m)x}\;+ O(\lambda)\\\nonumber
\end{eqnarray}
where $z=\rmi(p^*_n-p_m)/(2\sqrt{\lambda})$ .

In the limit $\lambda\rightarrow 0_+$ the terms $O(\lambda)$ vanish,
and for $\sqrt{\pi}\,z\,e^{z^2}\,{\rm erfc}(z)$ we have the result
(A.5). Thus we obtain (14) in this limit, which must be intended as
a prescription for the calculation of (13). We discuss this issue in
more detail in the Appendix.

Notice that for finite $\lambda>0$, the exact integrals over the
finite interval [0,\emph{L}] get rather involved and the one over
the resonant functions inside the barrier depends on the explicit
form of the functions other than on the position of the resonant
momenta.

The regularization works in a similar fashion for the crossed inner
products $\langle z_n|E\rangle$ and $\langle z_n|\phi_i\rangle$, as
well as for the symmetrical definition.

\section{Conclusions}

We have calculated some relevant inner products involving the
resonant eigenstates in the example of a 1-dimensional quantum
potential barrier with compact support. This old problem is
non-trivial since the modulus of the resonant solutions grows
exponentially at the spatial infinity, giving rise to infinite norms
and seemingly infinite inner products.

Among the variety of historical proposals to circumvent these
difficulties, we have adopted a Gaussian convergence factor, first
introduced by Zel'dovich, and carried out the limit of the integrals
where the factor fades off to unity. This prescription yields inner
products such that most of these states are orthogonal to each
other, except when they lie in a neighborhood defined by a
"divergence cone", in which case the product is infinite. Similarly,
scattering states (with real momentum $p>0$) and bound states (pure
imaginary momenta $\rmi q_i \;$, with $q_i>0$) are orthogonal to a
given resonant state $|z_n\rangle$ with momentum $p_n$ , except when
$p$ or $\rmi q_i$\hglue0.2cm lie in a neighborhood of $p_n$. Thus,
the resonant states share properties of the continuum and discrete
real spectra, namely (partial) orthogonality and the infinite norm
characteristic to the Dirac states.

This result is different from the full bi-orthogonality obtained by
the prescription of analytically continuing the finite integrals
from the upper complex momentum plane to the whole lower half plane,
where the resonant momenta lie but where the integrals are formally
divergent. Our limiting procedure instead extends the finite result
to the more modest $\pi/4$ angular sectors of the lower half plane
close to the real axis. We differ also from earlier attempts
\cite{Berggren1} using the Gaussian convergence factor, actually
limited to the products $\langle z_{-n}|z_m\rangle$ (where
$|z_{-n}\rangle$ was mistaken for the incoming state $|\tilde
z_n\rangle$), in that our exact calculation yields also $\langle
z_{-n}|z_n\rangle=0$ instead of a finite quantity normalizable to
$1$. It is interesting also the comparison with the results obtained
in the Friedrichs model \cite{Prigogine}\cite{Civitarese}.

Therefore the orthogonality and normalization properties obtained
depend on the prescription adopted, although one should expect, as a
signature of their consistency, that all of them lead to the same
unique result. For instance, the inner self-product $\langle
z_n|z_n\rangle$ yielding the square of the conventional norm
$\|z_n\|$, which is manifestly divergent on the same footing of the
scattering states, is finite for some prescriptions, whereas ours
recovers the infinite result.

We do not investigate here the question if a regularization by
different convergence factors (using for instance $e^{-\lambda
|x|^{\nu}}$ for some real value $1<\nu<2$) would yield narrower
divergence cones, thus approaching the result $\langle
z_n|z_m\rangle=0\;(n\neq m)$ which would seem more natural and
closer to the result of the analytical continuation prescription. On
the other hand, such a continuation in the momentum of the solutions
is a non-trivial matter \cite{Madrid2} related to the two-sheet
structure of the complex energy plane.

The completeness of the scattering (plus bound, if any) solutions in
Hilbert space is not inherited by the sole resonant (plus bound)
states in resonance expansions of the unity,  and a continuum of
complex energy "background" states must be included \cite{Berggren1}
\cite{Berggren2}\cite{Muga}. Any consistent unambiguous prescription
for the calculation of the inner products involving resonances
should be tested against the requirement of idempotence of the
unity, as discussed in the Introduction. It is not clear how could
it happen in the analytical continuation prescriptions, where the
claimed orthogonality between resonances (or between resonant and
anti-resonant states) is of the Kronecker-$\delta$ type, and the
cross inner products between bound, resonant and background states
have not been worked out.

We instead have orthogonality between most of the states and
divergent inner products between "neighbor" resonant (and continuum
and bound) states, which then have also infinite self-products
consistently with the well known infinite norm of the resonances. In
a general expansion involving bound, resonant (in variable number,
according to the poles encircled by the deformed integration path),
scattering and continuum background states, showing directly the
idempotence of the unity requires tackling the products involving
all of these families of states. In this paper we have done part of
this program and developed some regularization and calculational
techniques. Work is in progress on the inner products involving the
resonant and the background states, aiming to the eventual
cancelation of the divergences encountered.

\ack{Work supported by MEC project FIS2008-05705. The authors are
indebted to J. Le\'on for suggestions and helpful discussions. J.
Julve acknowledges the hospitality of the Dipartimento di Fisica
dell'Universit\`a di Bologna, Italy, where part of this work was
done.}

\appendix

\section{Limits of erfc({\it z})-related integrals}

We rely on the basic integral
\begin{equation}
J(k,\lambda)\equiv\int^{+\infty}_0 \rmd x\; \rme^{-\lambda
x^2}\rme^{\rmi kx }=\frac{\rmi}{k}\;\sqrt{\pi}\,z\,\rme^{z^2} {\rm
erfc}(z)\hskip 1.0cm (\lambda\;{\rm real}>0)
\end{equation}
which is directly related to (7.1.2) in \cite{Stegun}, where
$z=-\rmi k/(2\sqrt{\lambda})$ , and hence $k$ , can take any complex
value. Notice that on completing a square in the exponent and
changing the integration variable to $t=\sqrt\lambda\;x+z$ one
obtains the intermediate expression
\begin{equation}
J(k,\lambda)= \frac{1}{\sqrt\lambda}\;\rme^{z^2}\int^{\infty+z}_z
\rmd t\; \rme^{-t^2}
\end{equation}

The integral representation (7.1.2) is convergent when the path of
the complex integration variable $t$ approaches $\infty\;$ along a
direction $-\frac{\pi}{4}<{\rm arg}(t)<\frac{\pi}{4}$ . This
condition is fulfilled by (A.2) as long as $z$ remains finite.
However, in (A.1) not always the limit $\lambda\rightarrow 0$ (and
hence $z\rightarrow\infty$) can be taken before the integration is
carried out, because the function
$f_{\lambda}(x)\equiv\rme^{-\lambda x^2}$ does not converge
uniformly to the function $f_0(x)=1$ when $\lambda\rightarrow 0$ .

For Im $k\;>0$ the limit can be taken in the integrand in (A.1)
because the finiteness of the integral is always assured by the
convergence factor $\rme^{-({\rm Im}\,k)\,x}$, trivially yielding
$\rmi\,k^{-1}$. In this case $J(k,\lambda)$ converges to
$J(k,0)=\rmi k^{-1}$ when $\lambda\rightarrow 0$ .

For real $k$ the result
\begin{equation}
\int^{\infty}_0\rmd x \, e^{\rmi kx}= \rmi\,PV\frac{1}{k}+\pi
\,\delta(k)
\end{equation}
is also well known. Notice that it relies on adding to $k$ a small
imaginary part $\rmi\epsilon$, which still guarantees the
convergence when $\lambda\rightarrow 0$, but later in the limit
$\epsilon\rightarrow 0_+$ the result must be interpreted as a
distribution.

For Im $k\;<0$ the integration and the limit $\lambda\rightarrow 0$
do not commute. In that case we adopt the limit of the integral as a
prescription. In the right-hand side of (A.1), the limit
$\lambda\rightarrow 0$ can be directly inferred from the asymptotic
expansion
\begin{equation}
\fl\sqrt{\pi}\,z\,e^{z^2}\,{\rm
erfc}(z)\sim1+\sum_{m=1}^{\infty}(-1)^m\frac{1\cdot3\,...\,(2m-1)}{(2z^2)^m}
\;\;\;\;\;\;\;\;\;(\,z\rightarrow\infty \;,\;|{\rm arg}
z|<\frac{3\pi}{4}\,)
\end{equation}
(see (7.1.23) in \cite{Stegun}), namely
\begin{equation}
{\rm lim}_{z\rightarrow\infty}\sqrt{\pi}\,z\,e^{z^2}\,{\rm erfc}(z)=
\left\{
\begin{array}{ll}
1&\quad ,\quad -3\frac{\pi}{4}< {\rm arg}(z)<3\frac{\pi}{4}\\
\infty & \quad ,\quad {\rm otherwise}
\end{array} \right.
\end{equation}
which can also be numerically checked. In this way (A.5) yields an
extension of $J(k,0)$ to a new region of the lower half complex
plane:
\begin{equation}
 J(k,0)=\left\{
\begin{array}{ll}
\frac{\rmi}{k}&\quad ,\quad
-\frac{\pi}{4}<{\rm arg}(k)<5\frac{\pi}{4} \;\;\;,\;\; k\neq 0\\
\infty & \quad , \quad{\rm otherwise}
\end{array}
\right.
\end{equation}

\end{document}